\documentclass[sigconf,nonacm]{acmart}
\usepackage{amsmath,booktabs,graphicx}
\setcopyright{none}
\begin{document}
\title{SkillContrast: Difference-Guided Text Selection for Agent Skill Reranking}
\author{Jiandong Ding}
\affiliation{%
  \institution{Huawei Technologies Co., Ltd.}
  \city{Shanghai}
  \country{China}}
\email{dingjiandong2@huawei.com}

\author{Honglei Ji}
\affiliation{%
  \department{Department of Obstetrics}
  \institution{Shanghai East Hospital, Tongji University School of Medicine}
  \city{Shanghai}
  \postcode{200123}
  \country{China}}
\email{hongleijish@163.com}

\author{Ming Liu}
\affiliation{%
  \department{Department of Obstetrics}
  \institution{Shanghai East Hospital, Tongji University School of Medicine}
  \city{Shanghai}
  \postcode{200123}
  \country{China}}
\email{lmliuming@tongji.edu.cn}

\author{Tao Duan}
\affiliation{%
  \department{Department of Obstetrics}
  \institution{Shanghai East Hospital, Tongji University School of Medicine}
  \city{Shanghai}
  \postcode{200123}
  \country{China}}
\email{drduantao@126.com}
\renewcommand{\shortauthors}{Ding et al.}
\begin{abstract}
Similar agent skills can share instructions but differ in their conditions of use. Query-based text selection may retain shared instructions and omit these distinctions. We introduce \textbf{SkillContrast}, a training-free selector that compares retrieved skills and retains their differing text with local context for a pretrained reranker. On 1,235 requests from SameCapRisk-Bench, it yields 54--72 more clean hits---requests that retrieve a helpful skill without its marked risky sibling---than TF--IDF query selection at identical per-candidate input lengths, across 2 retrievers and 2 reranker sizes. Length-matched component replacements identify differing text as the main contributor in the primary setting, with smaller, mixed context effects. Relative to full skill bodies, SkillContrast uses 51.1--58.8\% fewer model-input tokens, with 10--18 fewer clean hits at 0.6B and matching or higher observed clean-hit counts at 4B. Candidate-relative differences thus complement query relevance in selecting compact reranking inputs.
\end{abstract}
\maketitle

\section{Introduction}
Agent systems retrieve reusable skills from libraries hosted on GitHub and agent platforms~\cite{skillret2026}. Several skills may serve the same purpose yet require different resources, preconditions, or input formats. Two migration skills, for example, may describe the same procedure but target different software versions. Both can look relevant to a request, although only one meets its conditions. We study how to make these distinctions visible to a reranker without passing it every candidate's full instructions.

Reading full skill bodies exposes information missing from short descriptions~\cite{router2026}. To shorten the reranker's input, a selector can instead choose passages---excerpts such as paragraphs or code blocks---by their query relevance~\cite{li2023keyb}. Yet similar candidates may share the most query-relevant instructions while differing in a prerequisite elsewhere. We therefore ask: can comparing the candidates reveal useful reranking text that query relevance alone overlooks?

We introduce \textbf{SkillContrast}, which compares textually similar retrieved skills and retains the blocks where their own instructions differ, together with opening and preceding context. A pretrained reranker then scores each candidate's selected text against the request, without further training. Skills without a close neighbor use query-based selection. We use maximal marginal relevance (MMR)~\cite{carbonell1998mmr} to balance these scores against redundancy when selecting the final list. Comparison chooses the text; the reranker judges its relevance to the request.

We evaluate on SameCapRisk-Bench~\cite{ding2026samecaprisk}, which tests whether top-3 results include a helpful skill without its marked risky sibling. This complements SkillRet's large-library retrieval evaluation~\cite{skillret2026} and R3-Skill's study of joint compatibility~\cite{r3skill2026}. Our matched-length comparisons show more clean hits with candidate differences than with TF--IDF query selection. Primary-setting replacements locate the main contribution in the differing text. Code and frozen-score table replay are available at \url{https://anonymous.4open.science/r/SkillContrast-92C8/}.

\section{Related Work}
\textbf{Agent skill retrieval.} Voyager stores executable skills for retrieval and reuse~\cite{wang2024voyager}, while ToolLLM retrieves APIs from a large tool catalog~\cite{qin2024toolllm}. These systems select reusable capabilities; our focus is which text from retrieved skill instructions reaches the reranker. Body-aware routing~\cite{router2026} and task-specific retrievers~\cite{skillret2026} improve query--skill matching at library scale. Among plausible results, shared descriptions can still obscure suitability. This motivates query-conditioned compatibility supervision in R3-Skill~\cite{r3skill2026} and generic-content score calibration in SkillSight~\cite{skillsight2026}. SkillContrast instead compares similar candidates to select their differing text for a frozen reranker. This exposes candidate distinctions without changing the retriever or training a new scoring model.

\textbf{Text selection for ranking.} KeyB~\cite{li2023keyb} selects query-relevant blocks using lexical or learned scores, while Provence~\cite{chirkova2025provence} prunes query-dependent sentences jointly with reranking. For similar skills, such selection may preserve shared instructions rather than distinguishing conditions. We compare equal-length, training-free selectors to isolate the effect of candidate comparison. List redundancy is a separate problem addressed by MMR~\cite{carbonell1998mmr} and its probabilistic batch extension SMMR~\cite{liakhnovich2025smmr}. We keep ordinary MMR fixed to isolate the selected text.

\section{Method}
\label{sec:method}
We ask whether differences improve selection at equal input length (RQ1), whether the gain comes from differences or context (RQ2), and how much reranking quality compact inputs retain (RQ3).

Given a request $q$, let $C=(c_1,\ldots,c_{20})$ be a fixed retrieved list and $d_i$ the visible body of candidate $c_i$. We select excerpts from $d_i$ to form the reranker input $d'_i$. All variants can access the same full bodies; they differ in the text passed to the neural reranker.

\begin{figure*}[t]
  \centering
  \includegraphics[width=\textwidth]{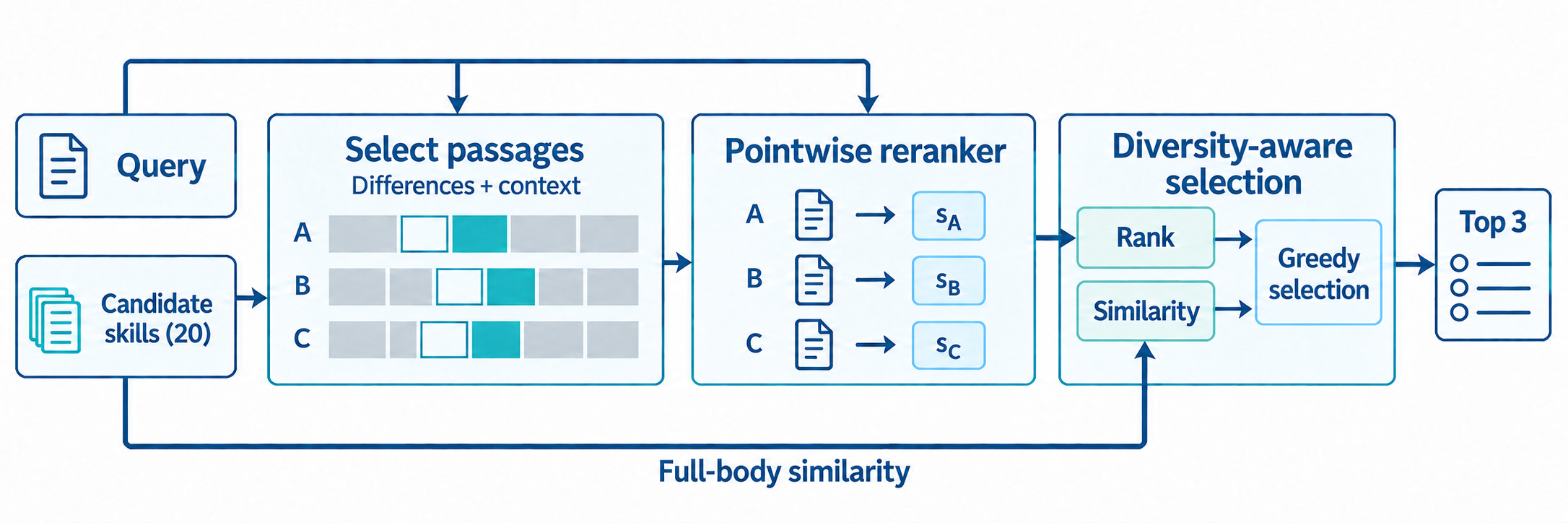}
  \caption{SkillContrast in a reranking pipeline. Our text-selection stage (``Select passages'') retains differences (teal), preceding context (outlined), and opening context. The frozen reranker and MMR are existing components. Full-body similarity feeds MMR separately; the query-based no-neighbor branch is not shown.}
  \Description{The query branches into passage selection and pointwise reranking. Candidate skills feed passage selection and a full-body-similarity path. Relevance ranks and similarity enter greedy selection, which returns 3 candidates.}
  \label{fig:method}
\end{figure*}

\paragraph{Selecting candidate differences.}
We represent full skill bodies with TF--IDF~\cite{salton1988termweighting}, connect candidates whose cosine similarity reaches 0.65, and compare them within connected components. TF--IDF uses lowercase unigrams and bigrams, with at most 160,000 features; it is fitted separately on each skill library. Grouping uses no benchmark family labels. Within each group, we align bodies line by line and map differences to paragraph and heading blocks, keeping fenced code intact. An insertion present only in another candidate selects adjacent text from the current candidate, never text copied across skills.

\begin{samepage}
For a candidate $c_i$ with comparison partners, let $\Gamma_i$ contain the indices of all other candidates in its connected component, including indirectly linked candidates, and let $\Delta_{ij}$ index blocks in $d_i$ selected by alignment with $d_j$. Define preceding blocks as $P(D)=\{b-1:b\in D,\ b>0\}$. The retained blocks $A_i$ and difference blocks $D_i$ are
\begin{equation}
\label{eq:passages}
A_i=\{0\}\cup D_i\cup P(D_i),\quad D_i=\bigcup_{j\in\Gamma_i}\Delta_{ij}.
\end{equation}
\end{samepage}
We retain $A_i$ in source order, counting overlaps once; block 0 provides the opening context. We do not add other text just to reach the target length. If no difference is found, the selected text is empty, or it exceeds a 768-token body target, we use the full body. This target governs text selection rather than imposing a maximum model-input length.

For a candidate with no comparison partner, we score heading, line, and sentence units by query--unit TF--IDF, protecting fenced and inline code. We keep the first unit as context. We then add positive-scoring units together with their predecessors, up to a 256-token body target. Ties follow source order. Short bodies and oversized first units use the full body. We also return the full body if no positive match fits or the exact token-count check detects overflow.

For candidates with usable differences, we tokenize the full input once and retain the tokens covering the selected blocks and their separators. Keeping their original order and token IDs avoids changes from decoding and retokenizing. The query, instruction, and scoring suffix stay unchanged, and selection uses no helpful/risky labels. Extraction and similarity still read full bodies; only the neural reranker's input is shortened (Figure~\ref{fig:method}).

\paragraph{Scoring and selecting skills.}
We next rank the selected inputs using frozen Qwen3-Reranker models~\cite{zhang2025qwen3}, scoring each $(q,d'_i)$ independently. For rank $\rho_i\in\{1,\ldots,N\}$ with deterministic ties, let $r_i=(N-\rho_i)/(N-1)$. MMR balances this relevance against similarity to selected candidates. Given the indices $S$ of selected candidates, MMR chooses
\begin{equation}
\label{eq:mmr}
\arg\max_{i\notin S}\;\lambda r_i-(1-\lambda)\max_{j\in S}G_{ij},
\end{equation}
where $G$ is full-body TF--IDF cosine similarity, the maximum is 0 when $S$ is empty, and ties favor the smaller $\rho_i$. We fix $N=20$, $K=3$, and $\lambda=0.75$.

With no selected candidate, the redundancy penalty is 0, so MMR chooses rank 1 first. It can diversify later choices but cannot remove a top-ranked risky skill. Such top-rank applicability errors must be resolved by text selection and reranking before MMR.

\begin{table*}[t]
\centering
\caption{Retrieval comparison on 1,235 requests. All methods use MMR(0.75) on their own top-20 candidates; SkillContrast reranks SkillSight candidates with 0.6B. Bold: best overall; underlined: best reference.}
\label{tab:systems}
\small
\begin{tabular*}{\textwidth}{@{\extracolsep{\fill}}lccccccccc}
\toprule
& \multicolumn{4}{c}{Generic retrieval controls} & \multicolumn{4}{c}{Public skill retrieval methods} & \textbf{Ours} \\
\cmidrule(lr){2-5}\cmidrule(lr){6-9}\cmidrule(l){10-10}
& BGE-M3 & \shortstack{BGE-\\reranker} & \shortstack{Qwen-\\listwise} & RRF & SkillRouter & SkillRet & R3-Skill & SkillSight & \textbf{SkillContrast} \\
\midrule
R@3 $\uparrow$ & 0.718 & 0.750 & 0.764 & 0.744 & 0.829 & 0.801 & \underline{0.846} & 0.837 & \textbf{0.867} \\
HSR@3 $\downarrow$ & 0.172 & 0.155 & 0.163 & 0.228 & 0.165 & 0.150 & \underline{0.137} & 0.165 & \textbf{0.113} \\
CH@3 $\uparrow$ & 0.657 & 0.721 & 0.704 & 0.632 & 0.777 & 0.758 & \underline{0.813} & 0.767 & \textbf{0.837} \\
\bottomrule
\end{tabular*}
\end{table*}

\begin{table*}[t]
\centering
\caption{Text selection on 1,235 requests. Each group fixes candidates, reranker, and MMR(0.75); compact strategies match length. All metrics use $K=3$. Bold: best, including ties. Token row: rounded means per candidate, full $\to$ compact; reductions use exact totals.}
\label{tab:compact_matched}
\small
\setlength{\tabcolsep}{3pt}
\begin{tabular*}{\textwidth}{@{\extracolsep{\fill}}lrrrrrrrrrrrr}
\toprule
& \multicolumn{3}{c}{SkillSight / 0.6B} & \multicolumn{3}{c}{SkillSight / 4B} & \multicolumn{3}{c}{SkillRet / 0.6B} & \multicolumn{3}{c}{SkillRet / 4B} \\
\cmidrule(lr){2-4}\cmidrule(lr){5-7}\cmidrule(lr){8-10}\cmidrule(lr){11-13}
Selection strategy & R $\uparrow$ & HSR $\downarrow$ & CH $\uparrow$ & R $\uparrow$ & HSR $\downarrow$ & CH $\uparrow$ & R $\uparrow$ & HSR $\downarrow$ & CH $\uparrow$ & R $\uparrow$ & HSR $\downarrow$ & CH $\uparrow$ \\
\midrule
Full body & \textbf{0.888} & \textbf{0.105} & \textbf{0.852} & 0.915 & 0.073 & 0.898 & \textbf{0.870} & 0.100 & \textbf{0.837} & \textbf{0.901} & 0.063 & \textbf{0.886} \\
\midrule
Prefix & 0.828 & 0.156 & 0.784 & 0.840 & 0.135 & 0.821 & 0.791 & 0.165 & 0.753 & 0.811 & 0.134 & 0.796 \\
Query-selected & 0.829 & 0.151 & 0.794 & 0.877 & 0.103 & 0.860 & 0.803 & 0.155 & 0.773 & 0.842 & 0.113 & 0.828 \\
SkillContrast (ours) & 0.867 & 0.113 & 0.837 & \textbf{0.923} & \textbf{0.061} & 0.906 & 0.856 & \textbf{0.099} & 0.829 & 0.900 & \textbf{0.057} & \textbf{0.886} \\
\midrule
Query-context variant & 0.871 & 0.113 & 0.840 & 0.922 & 0.062 & \textbf{0.908} & 0.854 & 0.105 & 0.826 & 0.899 & 0.061 & 0.884 \\
\midrule
Mean input tokens & \multicolumn{3}{c}{786 $\to$ 323 ($-$58.8\%)} & \multicolumn{3}{c}{786 $\to$ 323 ($-$58.8\%)} & \multicolumn{3}{c}{620 $\to$ 303 ($-$51.1\%)} & \multicolumn{3}{c}{620 $\to$ 303 ($-$51.1\%)} \\
\bottomrule
\end{tabular*}
\end{table*}

\begin{table}[t]
\centering
\caption{Component contributions at matched input length (SkillSight, 0.6B, MMR(0.75), 1,235 requests). Top-3 hit counts; $\Delta$Clean is relative to SkillContrast (dash: reference). BM25 is a separate selector control.}
\label{tab:components}
\small
\setlength{\tabcolsep}{3pt}
\begin{tabular*}{\columnwidth}{@{\extracolsep{\fill}}lrrrr}
\toprule
Input variant & Helpful $\uparrow$ & Risky $\downarrow$ & Clean $\uparrow$ & $\Delta$Clean \\
\midrule
SkillContrast (ours) & 1,071 & 139 & 1,034 & --- \\
\midrule
Replace differences & 995 & 203 & 956 & $-78$ \\
Replace first block & 1,074 & 133 & 1,034 & $0$ \\
Replace preceding context & 1,076 & 139 & 1,038 & $+4$ \\
\midrule
BM25-selected & 1,000 & 225 & 944 & $-90$ \\
\bottomrule
\end{tabular*}
\end{table}

\section{Experimental Setup}
We evaluate on SameCapRisk-Bench's 890-unit release. Its 823 main units produce 1,235 requests: 411 operation-contract, 774 source-role, and 50 caller-interface. The 67 development units produce 79 requests, excluded from main results. The reproduction bundle records the exact data identity.

We follow the benchmark's top-3 evaluation metrics. For each request, let $H$ indicate a hit on the marked helpful skill and $B$ a hit on its marked risky sibling. We report $\mathrm{R@3}=\mathbb{E}[H]$, harmful sibling rate $\mathrm{HSR@3}=\mathbb{E}[B]$, and CleanHit $\mathrm{CH@3}=\mathbb{E}[H(1-B)]$. A clean hit retrieves the helpful skill without its risky sibling. These metrics measure candidate exposure rather than execution failures.

We first compare retrieval pipelines using each baseline's own top-20 pool and MMR(0.75). To ask whether candidate differences improve text selection itself, we then use a $2\times2$ design: SkillSight and SkillRet candidates, each scored by Qwen3-Reranker-0.6B and 4B. Candidates, full-body similarities, reranker, and MMR stay fixed within each combination. Both sizes use bfloat16, SDPA, the same relevance instruction, and microbatch 1. The primary setting for component and parameter analyses is SkillSight with 0.6B; selector settings stay the same in the other 3 combinations.

For candidates with usable differences, controls match SkillContrast's exact input length. \emph{Prefix} takes the first body tokens; \emph{Query-selected} ranks units by query--unit TF--IDF and packs the same number. A separate \emph{Query-context variant} replaces preceding-only context while keeping differences and the first block. To isolate content choice, all compact strategies share the no-neighbor branch and full-body fallback, without padding. Packing preserves fenced code but may end optional prose at a token boundary. If a replacement cannot fit, we retain the original input.

To identify useful content, we replace differences, the opening block, or preceding context with equally long query-relevant text. We change only tokens belonging solely to that component; shared tokens remain, so replacing context cannot remove a difference. The BM25-selected control uses BM25 scoring~\cite{robertson2009bm25}, adapting KeyB's lexical selection~\cite{li2023keyb} to our source units, input budgets, and reranker. Paired intervals use 10,000 bootstrap draws stratified by protocol over 410 benchmark groups (375 operation-contract, 10 source-role, 25 caller-interface), keeping dependent requests together. We report exploratory 95\% intervals and Bonferroni-adjusted 98.75\% intervals for the 4 primary CleanHit comparisons; adjusted intervals have limited tail precision.

\section{Results and Analysis}
Table~\ref{tab:systems} compares full retrieval pipelines using MMR(0.75). Baselines use their top-20 lists recorded in the benchmark, while SkillContrast reranks SkillSight's list. Its CH@3 is 0.837 versus 0.813 for R3-Skill, the strongest reference. Because the retrieved candidates can differ, we fix them below to isolate text selection.

\paragraph{\textnormal{\textbf{RQ1: Do differences improve selection at equal length?}}}
With candidates, reranker, and MMR fixed (Table~\ref{tab:compact_matched}), SkillContrast improves helpful and clean hits and reduces risky exposures over both equal-length controls in all 4 settings. Against query selection, it adds 54 and 57 clean hits on SkillSight with 0.6B and 4B, and 69 and 72 on SkillRet; gains over prefixes range from 66 to 111. The corresponding adjusted CH@3 intervals are $[0.0198,0.0684]$, $[0.0257,0.0683]$, $[0.0309,0.0808]$, and $[0.0366,0.0810]$, all above 0 after correction for 4 comparisons. These gains reflect content choice, not input length.

Gains concentrate in operation-contract requests across all 4 settings. For example, SkillRet with 4B gains 72 clean hits in this protocol, versus $-2$ for source-role and $+2$ for caller-interface, yielding the overall $+72$.

\paragraph{\textnormal{\textbf{RQ2: Do gains come from differences or context?}}}
We next identify which content contributes to this advantage. Replacing difference-only tokens with equally long query-relevant text loses 78 clean hits in the primary setting (1,034 to 956; Table~\ref{tab:components}), with 76 fewer helpful hits and 64 more risky exposures. Retaining differences gives a 6.32-percentage-point CH@3 advantage (exploratory 95\% group-bootstrap interval: 4.47--8.16 points). Query-relevant replacement does not recover the removed information; the separate BM25 selector loses 90 clean hits.

In contrast, replacing the opening block or preceding context changes clean hits by 0 and $+4$ while retaining differences. Query-based preceding context gives $+4$ and $+2$ on SkillSight with 0.6B and 4B, but $-4$ and $-2$ on SkillRet. Differences drive the primary-setting gain; query-relevant context can complement them, with mixed effects across candidate pools.

The matched-length gains persist in local threshold and MMR scans. Clean hits nevertheless fall to 896 at $\lambda=0.85$ as redundancy receives less weight. The scans vary each parameter separately, keeping the text-selection targets fixed.

\begin{figure}[t]
  \centering
  \includegraphics[width=\columnwidth]{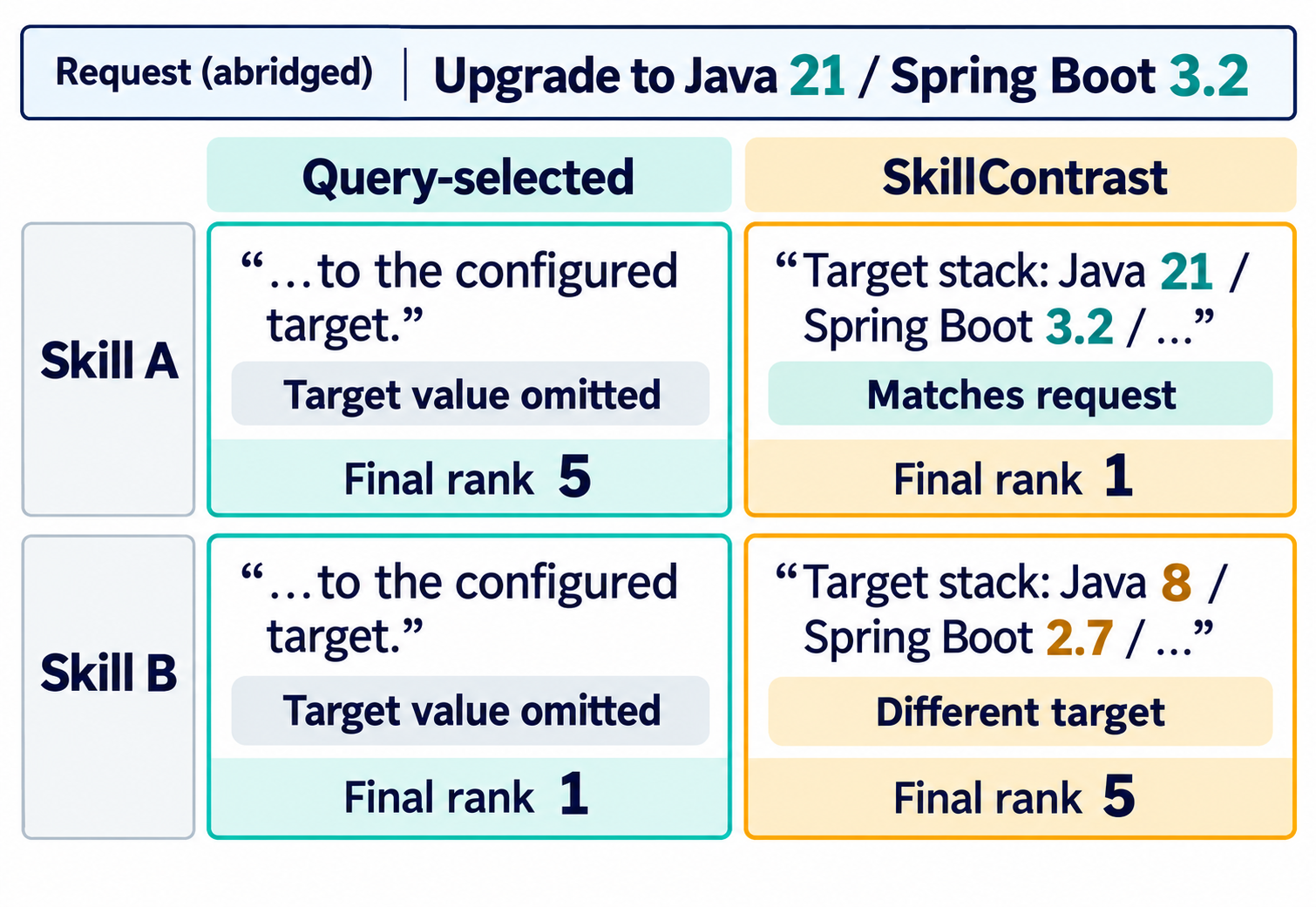}
  \caption{Two migration skills, 1 requested target (SkillSight, 0.6B). Excerpts are abridged; inputs match length per candidate (A: 328 tokens; B: 327). Final ranks use the same MMR over 20 candidates. Match labels are offline interpretations.}
  \Description{A matrix aligns 2 skills with 2 passage selectors. The request targets Java 21 and Spring Boot 3.2. Both query-selected excerpts refer to a configured target; SkillContrast exposes Java 21 and Spring Boot 3.2 for A, versus Java 8 and Spring Boot 2.7 for B. A moves from final rank 5 to 1, and B from 1 to 5.}
  \label{fig:case}
\end{figure}

\noindent\begingroup%
\setlength{\fboxsep}{3pt}%
\fcolorbox{black!35}{black!3}{%
\begin{minipage}{\dimexpr\columnwidth-2\fboxsep-2\fboxrule\relax}
\textbf{Case study: exposing the requested target.}\par
In Figure~\ref{fig:case}, query selection retains the shared migration procedure but omits the target versions. The excerpts remain relevant to the migration task without exposing the condition that distinguishes the two skills.
\par\setlength{\parskip}{3pt}\noindent
SkillContrast exposes these target configurations. With the reranker and per-candidate input lengths held fixed, the matching skill moves to the top of the final list. This selected example illustrates how candidate differences can make applicability conditions visible during reranking.
\end{minipage}}\par
\endgroup

\paragraph{\textnormal{\textbf{RQ3: How much quality do compact inputs retain?}}}
We now compare compact inputs with full bodies. The token row in Table~\ref{tab:compact_matched} reports 58.8\% less input on SkillSight and 51.1\% on SkillRet at both model sizes; compact controls have identical totals. Relative to full bodies, SkillContrast loses 18 and 10 clean hits at 0.6B. At 4B, it gains 10 on SkillSight and matches SkillRet's 1,094, with 15 and 8 fewer risky exposures; SkillRet loses 1 helpful hit.

Shorter model inputs do not guarantee a faster pipeline. We timed a single pass with the 0.6B reranker on the original benchmark data, rotating strategy order per request. Relative to full bodies, SkillContrast reduced online time by 10.4\%, saving 76.8\,ms per request. Timing covered input construction, encoding, 20 model forwards, and MMR. We compared 1,229 requests valid across strategies. Input validation rejected 8 of 3,705 records. Index and source-feature preparation cost 25.1\,s and 135.0\,s separately. GPU processes were sampled every second, with no other CUDA process observed. This single-pass measurement, with CPU exclusivity unverified, separates preparation cost from online reranking time.

\section{Conclusion}
SkillContrast retains candidate differences with local context for a pretrained reranker. Across 4 SameCapRisk-Bench settings, it adds 54--72 clean hits over TF--IDF query selection at matched lengths. Primary-setting replacements identify differing text as the main contributor. It cuts model-input tokens by 51.1--58.8\%, with 10--18 fewer clean hits at 0.6B and matching or higher observed counts at 4B.

Our evaluation uses skill text and benchmark annotations, without executing skills or testing agents on users. CleanHit excludes the marked risky sibling, not all unsafe candidates. Agent use still requires permission and execution checks. Future work should test independent skill libraries with labels for applicability. This would assess whether the advantage transfers beyond this benchmark.

\bibliographystyle{ACM-Reference-Format}
\bibliography{references}
\end{document}